*Structural bioinformatics*

# Daisy: An integrated repeat protein curation service

Manuel Bezerra Brandao[1,*], Ronaldo Tunque[1] and Layla Hirsh[1]

[1]Department of Engineering, Pontificia Universidad Católica del Perú, Lima, 32, Perú

*To whom correspondence should be addressed.

**Abstract**
Tandem repeats in proteins identification, classification and curation is a complex process that requires manual processing from experts, processing power and time. There are recent and relevant advances applying machine learning for protein structure prediction and repeat classification that are useful for this process. However, no service contemplates required databases and software to supplement researching on repeat proteins. In this publication we present Daisy, an integrated repeat protein curation web service. This service can process Protein Data Bank (PDB) and the AlphaFold Database entries for tandem repeats identification. In addition, it uses an algorithm to search a sequence against a library of Pfam hidden Markov model (HMM). Repeat classifications are associated with the identified families through RepeatsDB. This prediction is considered for enhancing the ReUPred algorithm execution and hastening the repeat units identification process. The service can also operate every associated PDB and AlphaFold structure with a UniProt proteome registry.
**Availability:** The Daisy web service is freely accessible at <u>daisy.bioinformatica.org</u>
**Contact:** mbezerrabrandao@pucp.edu.pe

## 1 Introduction

Repeat proteins have an increasing interest recently due to their stable structural folds, high evolutionary conservation and repertoire of functions provided by these proteins (Chakrabarty and Parekh 2022). It is also known that at least one out of three human proteins are considered to have tandem repeats (Jorda and Kajava 2010). The repeat units in proteins identification and classification is a complex process that requires manual processing from experts (Hirsh et al. 2016).

For the curation process, many bioinformatics web services are used. For example, it is necessary to access protein sequences and structures from different databases (Berman et al. 2000; Di Domenico et al. 2014; The UniProt Consortium 2019). In addition, there are new services that could make this curation process more efficient, like repeat protein classification predictors (Muroya Tokushima and Hirsh 2022; Tenorio Ku and Hirsh 2021) protein structure prediction methods (Jumper et al. 2021; Palomino and Hirsh 2021; Tunyasuvunakool et al. 2021) and tandem repeat in proteins identification software (Hirsh et al. 2016, 2017, 2018; Pedraza and Hirsh 2019).

Despite this type of diverse bioinformatics software that exists and is available for any user, to our knowledge there is no web service for repeat proteins curation that gathers all the information needed. This process, without an integrated tool, is more complex and cumbersome, considering that there are a lot of repeat proteins to characterize and annotate (Chakrabarty and Parekh 2022; Di Domenico et al. 2014).

To improve this situation, this research presents the first version of the Daisy software, an integrated web service for repeat protein curation. This service has been developed considering the possibility of working with known protein structures or prediction based on a protein sequence. Also, it improves the efficiency of a tandem repeats identification algorithm by considering a repeat classification prediction using a protein functional analysis algorithm. In addition, the web user interface was designed taking following usability guidelines for bioinformatics web services (Bezerra Brandao, Hirsh, and Pow Sang 2021) to ensure the ease of use of this software.

## 2 Components

The presented service uses an updated classification of the 3D structures of proteins with repeats. Specifically, subclasses belonging to the class III (Elongated structures where repetitive units require one another to maintain structure), class IV ("Closed" structures where repetitive units need one another to have structure) and class V ("Beads on a string" structures which repeats are large enough to fold independently) are the ones considered for the repeat protein curation process (Kajava 2012).

To develop this service, it was needed to select bioinformatics tools and databases that researchers use for the tandem repeats in proteins identification process. Each component of the integrated service is described below.

## 2.1 Protein structure databases

The curation process for this service starts with obtaining the selected protein structure files. Initially, the Protein Data Bank (Berman et al. 2000) is the ideal database to obtain known protein structure files. Nevertheless, there are current relevant advances in deep learning models applied for protein structure prediction (Tunyasuvunakool et al. 2021). Therefore, it was also considered for this service to be integrated into the AlphaFold database (Jumper et al. 2021; Varadi et al. 2022).

This integrated web service can obtain the structure files of a protein through it's accession identification by connecting to the databases. To process a known protein structure, it will only require it's PDB accession id. Otherwise, to work with the predicted structure of a protein sequence, the service will need the AlphaFold registry id. This corresponds to the UniProt sequence accession id.

For this first component of the curation process, the Daisy service allows to visualize the protein structure files based on the accession id (Fig 1). Once the structure files are obtained, the service executes the next component.

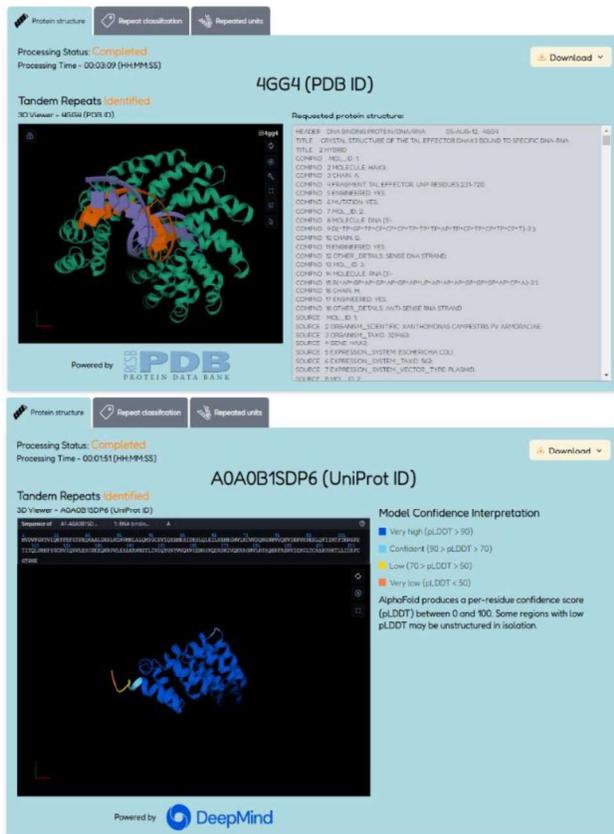

**Fig. 1** Protein structure visualization of the Daisy web service. The first part of the figure shows the visualization of a known protein structure obtained at the PDB database (Berman et al. 2000). The last part of the figure shows the visualization of a structure prediction obtained at the AlphaFold database (Jumper et al. 2021; Varadi et al. 2022).

## 2.2 Repeat protein classification prediction

The tandem repeats identification process on this service continues by predicting possible repeat classification for the evaluated protein structure. For this task, PfamScan, a protein functional analysis algorithm, was selected (Li et al. 2015; Mistry, Bateman, and Finn 2007). It searches a sequence against a library of Pfam (Mistry et al. 2021) hidden Markov model (HMM). This algorithm allows to obtain possible protein families for a sequence file (.fasta). In addition, RepeatsDB (Di Domenico et al. 2014) contains known repeat classifications for 776 protein families. Each of these families can have one or more subclasses associated.

For this step of the curation process, the Daisy software delivers each possible family for each protein structure chain through PfamScan (Fig 2). Also, each family contains a list of known repeat classifications associations obtained from RepeatsDB. This allows to elaborate a list of possible classifications for each chain of the evaluated structure. With the prediction results, the integrated service efficiently continues the repeat protein curation process, executing the last component for the probable subclasses list. If this list is empty, but there is at least one known repeat protein family, all possible classifications are considered for the process.

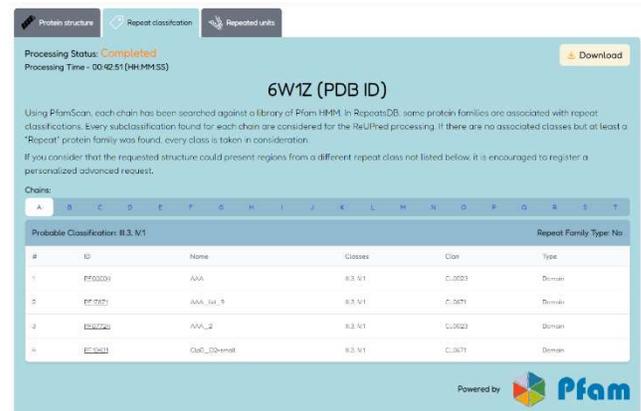

**Fig. 2** Repeat classification prediction of the Daisy web service. For each chain possible protein families obtained through PfamScan (Li et al. 2015; Mistry, Bateman, and Finn 2007) are shown. Each family presents a list of associated repeat classification obtained from RepeatsDB (Di Domenico et al. 2014). Each probable class is considered for the ReUPred processing.

## 2.3 Tandem repeats identification

For this last processing step, the ReUPred algorithm is optimized through the RepeatsDB-Lite software (Hirsh et al. 2018). This algorithm uses an iterative divide-and-conquer approach. Each iteration corresponds to a structural search, i.e., structural alignment of the query structure against all SRUL (structural repeat unit library). The predicted unit corresponds to the aligned region in the query. At each cycle, the algorithm forks (divide). Two new input structures are created, corresponding to the N- and C-terminal flanking fragments of the predicted unit and two new cycles (structural searches) are performed. After the "master" unit is found, it populates an ad hoc library, and all newly predicted units are included for search in the following cycles. The algorithm stops when the entire input protein is consumed. The predicted units are then collected and evaluated together (conquer). Suppose the result does not satisfy a set of rules. In that case, the structural alignment filters for the "master" unit are relaxed and the entire iterative part is repeated from the beginning for up to four increasingly relaxed iterations (Hirsh et al. 2016).

A valid solution for ReUPred is obtained when at least three units are found and their proximity in the sequence is ensured by at least one of two simple rules to measure unit proximity: (1) the total number of gaps between units is less than 40 residues, (2) the number of non-adjacent units divided by the total number of predicted units is less or equal to 0.25. For the algorithm, a unit must have at least 13 amino acids (Hirsh et al. 2016).

This RepeatsDB-Lite software processes the protein through a structure file (.pdb) and tries to identify every possible repeat region and units (Hirsh et al. 2018). It manages to identify tandem repeats in a protein, their classification and units' alignments (Hirsh et al. 2018). However, this algorithm requires a lot of processing power and time because it considers all possible subclasses inside classes III, IV and V.

To make it more efficient, the algorithm was modified to use a repeat class and subclass as input for the execution. In this way, the repeat classes whose probability reaches the required configurable threshold are the only ones considered in the software execution. This allows to require less processing power and time for the execution of the ReUPred algorithm for obtaining the desired outputs for each tandem repeat identified on the evaluated protein structure.

The Daisy web service shows the structural alignment of the identified repeated units (Fig 3). It also presents a visualizer for the repeated region's sequence and secondary structure alignments and the average root-mean-square deviation (RMSD). Finally, the user can download every output file, including each repeat unit structure and the similarity matrix.

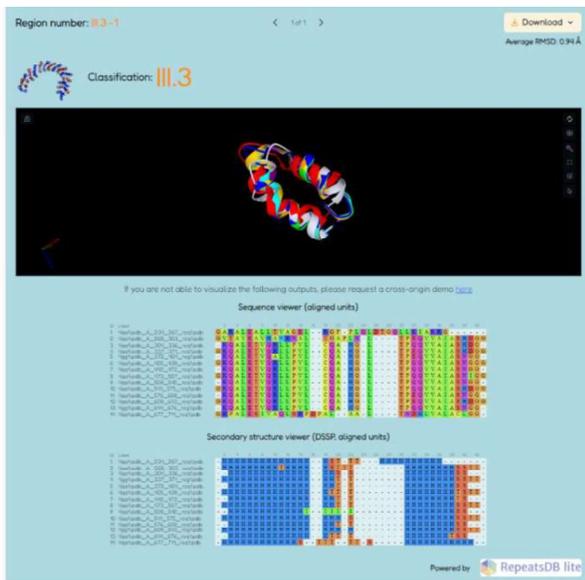

**Fig. 3** Tandem repeat regions visualization of the Daisy web service. Based on the RepeatsDB-Lite software, an optimization of the ReUPred algorithm identifies tandem repeats regions on the evaluated protein (Hirsh et al. 2018). The identified repeat class and subclass, the structural, sequence and secondary structure alignments can be visualized. Every output can also be downloaded by the user.

## 3   Web service

To be able to access this integrated software, a React app was developed. The user interface was designed taking in consideration bioinformatics web services usability guidelines (Bezerra Brandao, Hirsh, and Pow Sang 2021). It allows the submit of a request for the complete processing of a protein (Fig. 4). Users are asked to enter the corresponding protein accession id and their email. Once the request is submitted, a request number will be generated randomly for the user to retrieve the outputs at the end of the request.

After the process has finished, the user would be capable of observing and downloading the requested protein structure file. Also, each probable repeat classification can be downloaded for each protein chain through PfamScan and RepeatsDB. Lastly, all the outputs will be displayed if the modified ReUPred algorithm identifies any tandem repeats in the protein. Also, it is possible to download the results per region, including the repeat units' structures files, the aligned units' files and the correlation matrix.

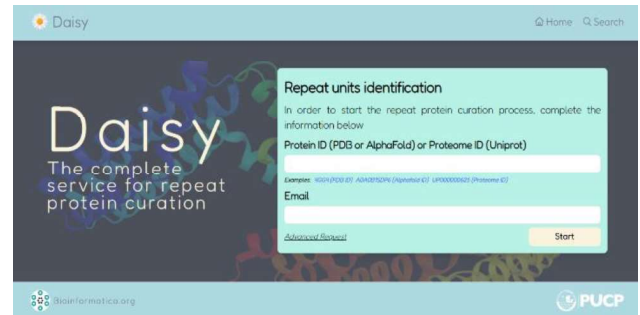

**Fig. 4** Homepage of the Daisy web service. Here the user can submit a processing request for any protein structure given an accession id on the PDB or AlphaFold databases.

In addition, the server allows to configure the execution parameters as an advanced request. The user can select specific subclasses to execute the ReUPred algorithm on the requested protein structure (Fig. 5).

**Fig. 5** Advanced request on the Daisy web service with selection of specific subclasses.

Lastly, the Daisy web service can process a requested proteome. For this kind of request, the server will search for each cross-referenced structure for each proteome component on the UniProt database (The UniProt Consortium 2019). The result for each structure can be visualized and downloaded by the user. In addition, the results can be ordered and filtered by structure database, tandem repeat regions presence, proteome component and execution time (Fig. 6).

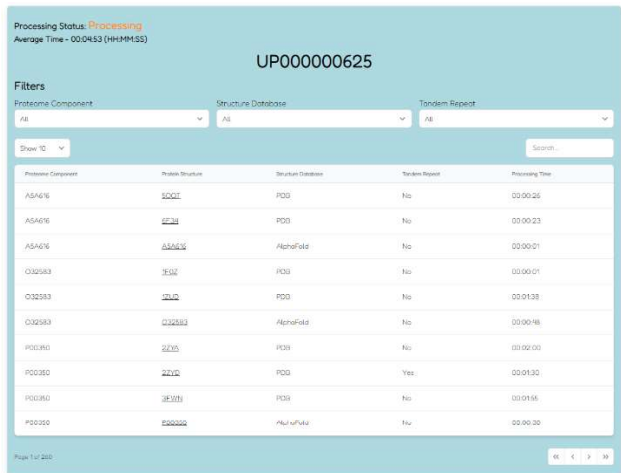

**Fig. 6** Proteome request results on the Daisy web service. Here the user can select a specific protein structure result. Also, the user can filter and order the results according to the structure database, tandem repeat regions, proteome component or execution time. In addition, an average processing time can be observed.

## 4   Discussion

A sample of the proteome Escherichia coli (strain K12) (UP000000625) was processed to evaluate the performance of the Daisy Web Service. Below is a summary of the obtained results (Table 1).

**Table. 1** Proteome processing results on the Daisy web service for the Escherichia coli (strain K12) (UP000000625). For the proteome, the number of processed protein structures, number of structures with tandem repeat regions (TDRR) and average processing time (APT) are presented.

| Proteome ID | UP000000625 |
|---|---|
| Processed structures | 2300 |
| Processed structures (PDB) | 2090 |
| Processed structures (AlphaFold) | 210 |
| APT (seconds) | 110.15 |
| Structures with TRR | 155 |
| APT for structures with TDRR (seconds) | 761.94 |
| APT for structures without TDRR (seconds) | 63.05 |
| APT for PDB structures with TDRR (seconds) | 797.33 |
| APT for AlphaFold structures with TDRR (seconds) | 454.46 |
| Average number of regions for structures with TDRR | 1.89 |
| APT for each TDRR | 539.44 |

It is observed that 2300 structures have processed, where 90.87% were obtained from the PDB and 9.13% from the AlphaFold database. This proteome's average processing time is approximately 1.84 minutes per structure. Likewise, the average processing time for an AlphaFold structure with tandem repeat regions is approximately 7.57 minutes. Meanwhile, the average processing time for a PDB structure with tandem repeat regions is 13.29 minutes. It is important to note that PDB structures tend to be significantly more extensive than most AlphaFold structures. The average execution time for a single chain in the RepeatsDB-Lite software is mentioned to be many minutes but varies depending on the class of the master unit (Hirsh et al. 2018)

It is also noted that the average processing time for structures without tandem repeat regions is approximately 1.05 minutes for the processed proteome. Finally, it is also observed that some structures present more than one identified repeat region on their chain. The average number of regions on structures with identified repeats is 1.89. This leads to the conclusion that the average processing time for each repeat region is approximately 8.99 minutes.

## 5   Conclusions

An integrated web service was designed and developed for the repeat protein curation process in this research. The Daisy web service is online and is capable of receiving requests for any known or predicted protein structure available at the PDB or AlphaFold databases, respectively (Berman et al. 2000; Varadi et al. 2022).

For the curation process, this service uses a protein functional analysis algorithm to predict the repeat classification of a solicited protein per chain. This prediction is a crucial input for an efficient execution of the optimized ReUPred algorithm. Every output is available for visualizing and downloading for each tandem repeat region identified. In addition, if a user considers that a structure may have a region of a repeated class that has not been contemplated, it is also possible to register an advanced request to perform a personalized execution. However, if no region is identified during the process, it does not necessarily imply that the evaluated structure is not considered to present tandem repeat regions. Many factors influence the identification and classification of tandem repeats and new annotations are constantly being made.

The future work for the Daisy web service consists of implementing each request's email notifications. Additionally, it is planned to be able to accept a user's local structure file input in the required formats in a future update. Finally, the reader is encouraged to use this integrated web service for repeat protein curation and the generated data.

**Acknowledgements**

European Union's Horizon 2020 research and innovation programme under the Marie Skłodowska-Curie, REFRACT [823886]. We render thanks to Luiggi Tenorio Ku for letting us use the deep learning model for repeat protein classification prediction he developed. We also acknowledge Emilio García Ríos and Corrado Dally Scaletti for helping deploy the Daisy web service on the Bioinformatica.org website.

**CRediT author statement**

**Manuel Bezerra Brandao**: Conceptualization, Software, Writing - Original Draft, Visualization **Ronaldo Tunque**: Software, Visualization **Layla Hirsh**: Conceptualization, Software, Data Curation, Validation, Resources, Supervision, Writing - Review & Editing

**Funding**

This work received no external funding.

*Conflict of Interest:* none declared.